# Solvents based model reduction of linear systems

## Karim Cherifi* and Kamel Hariche§

*Corresponding author, Institute of electrical and electronic engineering, Rue de l'independence, Boumerdes, 35000, Algeria, E-mail: cherifikarimdz@gmail.com

§Institute of electrical and electronic engineering, Rue de l'independence, Boumerdes, 35000, Algeria, E-mail: khariche@yahoo.com

**Abstract**

Model order reduction is the approximation of dynamical systems into equivalent systems with smaller order. Model reduction has been studied extensively for different types of systems. In this paper, we present two methods for multi input multi output linear systems. These methods are based on solvents, also called block poles. These methods are particularly suitable if the given system is in matrix transfer function form. The first method eliminates solvents one by one whereas, the second method can eliminate multiple solvents at the same time. The two presented methods are implemented in MATLAB in order to provide a systematic method for the model order reduction of MIMO linear systems.

# 1 Introduction

Model reduction has been extensively studied in recent years. This is due to the fact that nowadays we are dealing with large systems or switching systems that have a lot of configurations which require a lot of computation power and memory space. Depending on the application, computation capabilities are limited and the results are needed rapidly. For example, in real time embedded systems applications or when having to iterate thousands of simulations to tune some parameters for a system that is constantly changing. Then, it is required to make these systems simpler. Model reduction approximates the original system into a simpler system with smaller order. In other cases, the given system is unnecessarily cumbersome. Quite often not all the components make significant contribution to the system dynamic behavior. And even if modern computers are computationally capable of dealing with the original systems, it would not be energy or time efficient.

In this paper, we deal with linear time invariant systems either represented by a state space equation or a matrix transfer function G(s):

$$\dot{X} = AX + BU$$
$$Y = CX + DU \qquad (1)$$

$$G(s) = C(sI - A)^{-1}B + D \qquad (2)$$

A is of size $n \times n$, B of size $n \times m$., C of size $p \times n$ and D is of size $p \times m$. This system has order n. The goal of model reduction is to transform G(s) into an equivalent low dimensional system with far lower storage requirements and evaluation time. The order of the resulting system is $\bar{n}$ such that $\bar{n} < n$. The resulting reduced model is generally used to develop a low dimensional controller suitable for real time applications.

Many methods were proposed in the past [1-6] but few were based on matrix polynomials [7, 8]. In this paper we present two new methods based on solvents.

First, we start by reviewing the basic definitions of matrix polynomials and solvents. Then, we introduce the main model reduction techniques based on solvents. After that, the two contributed methods are presented successively. In order to illustrate the implementation of the two algorithms in MATLAB, an example is presented.

## 2 Matrix polynomials and solvents

In this section, some basic definitions of matrix polynomials are presented. The theory of matrix polynomial is by itself a field and will not be discussed extensively here. For more on this, The references [9, 10] can be consulted. In MIMO systems, matrix transfer functions are matrices where each entry is a transfer function. A MIMO transfer function can be represented as a fraction of polynomial matrices as in:

$$G(s) = N_R(s)D_R(s)^{-1} = D_L(s)^{-1}N_L(s) \qquad (3)$$

where

$N_R$ and $D_R$ which are matrix polynomials of orders q×p and p×p respectively form a right matrix fraction.

$N_L$ and $D_L$ which are matrix polynomials of orders q×q and q×p respectively form a left matrix fraction.

A matrix polynomial A(s) or $A(\lambda)$ is of the form:

$$A(\lambda) = A_0\lambda^r + A_1\lambda^{r-1} + ... + A_{r-1}\lambda + A_r \qquad (4)$$

A latent root $\lambda_i$ is the solution of the equation:

$$\det(A(\lambda)) = 0 \qquad (5)$$

The corresponding primary right latent vector p is the solution of

$$A(\lambda_i)p = o \qquad (6)$$

The right part of equation (6) is the zero vector with size corresponding to the size of $A_i$. The latent roots together with latent vectors form a latent structure. In a similar fashion, a left latent vector is the solution of the equation:

$$q^T A(\lambda_i) = o \qquad (7)$$

Solvents are important in matrix polynomials [11, 12] and are defined as follows:

*Definition 1: [12]*

*A right solvent or right block root is defined as the solution of:*

$$R^r + A_1 R^{r-1} + ... + A_{r-1}R + A_r = o \qquad (8)$$

*A left solvent or left block root is defined as the solution of:*

$$L^r + L^{r-1}A_1 + ... + LA_{r-1} + A_r = o \qquad (9)$$

*A matrix R (L) can be called a right (left) solvent if and only if $A(\lambda)$ in equation (4) can be divided exactly by $\lambda I - R$ or $\lambda I - L$ for a left solvent. If the given matrix X does not divide $A(\lambda)$ exactly, this will result in a reminder matrix $\Re$ for a right division or $\Gamma$ for a left division such that*

$$A(\lambda) = Q(\lambda)(\lambda I - X) + \Re \qquad (10)$$

$$A(\lambda) = (\lambda I - X)S(\lambda) + \Gamma \qquad (11)$$

A complete set of solvents is a special set of solvents that completely describes the latent structure of the matrix polynomial with the following properties:

$$\begin{cases} \cup \sigma(R_i) = \sigma(A_c) \\ \sigma(R_i) \cup \sigma(R_j) = \varnothing \\ det(V_R(R_1, R_2, ..., R_r)) \neq 0 \end{cases} \quad (12)$$

where $\sigma$ represents the spectrum of the matrix $A_c$ is in controller canonical form. $V_R$ is the block Vandermonde matrix in the form:

$$V_R = \begin{bmatrix} I_m & I_m & ... & I_m \\ R_1 & R_2 & ... & R_r \\ \vdots & \vdots & ... & \vdots \\ R_1^{r-1} & R_2^{r-1} & ... & R_r^{r-1} \end{bmatrix} \quad (13)$$

If $V_R$ is needed, the complete set of solvents has to be computed. The methods that can be used to compute the complete set of solvents are discussed in section 4. Solvents can also be computed individually using the eigenvalues and eigenvectors of the matrix $A_c$. This method is discussed more in depth in section 3.

Many techniques were presented in the past for model reduction [1-3, 13-18]. These techniques have tried to improve: storage, computational speed and accuracy. Most methods can be categorized into two main approaches: Krylov based subspaces and Truncation. Among the proposed methods, we cite: the Padé via lanczos method, the Arnoldi and Prima method, the Laguene method, the balanced truncation method, the optimal Hankel norm method and the Proper orthogonal decomposition (POD) method. Other algorithms are based on the factorization of transfer functions by eliminating poles (or solvents in case of MIMO systems).

All model reduction methods based on solvents [7, 8] require a complete set of solvents. The solvents are eliminated one by one until the threshold error is exceeded. In this paper we present two new methods. In section 3, the method presented does not require a full set of solvents. The solvents are computed one by one directly from eigenvalues and eigenvectors of the matrix A. On the other hand, in the method presented in section 4, multiple solvents are eliminated at the same time and only the solvents corresponding to the dominant poles are kept in the final model. These methods can be considered good if they fit these three criteria: computational efficiency of the algorithm, preservation of the properties of the original system (such as stability) and minimization of the approximation error.

## 3 Method based on eigenvalues

There are several methods to obtain solvents. One of these methods is using latent roots and corresponding latent vectors.

***Theorem 1:***

*If $A(\lambda)$ has n linearly independent right latent vectors $V = \{v_1, v_2, ..., v_n\}$ corresponding to latent roots $\Lambda = \{\lambda_1, \lambda_2, ..., \lambda_n\}$ then a right solvent R can be computed as:*

$$R = V \Lambda V^{-1} \qquad (14)$$

*The same can be concluded about left solvents using latent roots and their corresponding left latent vectors.*

Proof: [19].

Given a MIMO transfer function which can be written as a coprime fraction:

$$G(s) = N_R(s) D_R(s)^{-1} \qquad (15)$$

The idea here is to eliminate poles which have the weakest effect on the system response. Since latent roots correspond to poles of the system, all latent roots are computed; Afterwards, they are sorted to get the smallest ones (farthest left of the s-plane). These have the smallest effect on the system and if they are far enough from the poles which are the closest to the imaginary axis, then these can be eliminated without affecting the response of the system. Next, a solvent R is computed using the chosen latent roots and their corresponding latent vectors. Finally, this solvent can be eliminated by equation (10). In this case, since R is a solvent, it divides exactly $D_R$.

$$D_R(\lambda) = D_{MOR}(\lambda)(\lambda I - R) \qquad (16)$$

where $D_{MOR}$ is the reduced order denominator of the coprime fraction.

In order to keep the system proper, the nominator $N_R(s)$ is also reduced such that:

$$N_R(\lambda) = N_{MOR}(\lambda)(\lambda I - R) + \Re \qquad (17)$$

Notice that $(\lambda I - R)$ does not divide $N_R(\lambda)$ exactly. This means that the remainder $\Re$ is neglected. This is why this step is only performed when absolutely necessary.

In order to test the quality of the new model, an error criterion has to be used. In our methods we use an output error such that:

$$\max_{t>0} \left\| y(t) - \hat{y}(t) \right\|_\infty \leq \left\| G - \hat{G} \right\|_{H_2} \qquad (18)$$

The derivation of this inequality can be found in [20].
Practically, the $H_2$ norm can be computed as in (19).

$$\|F\|_2^2 = tr(B^T Q B) = tr(C P C^T) \tag{19}$$

Where P and Q are the controllability and observability matrices respectively.

The procedure to compute the reduced order model is as follow:
**INPUT:** transfer function $G(s)$
**OUTPUT:** reduced order $G_{mor}(s)$

***Step 1:*** Compute $G(s) = N_R(s) D_R(s)^{-1}$ and the latent roots $\Lambda = \{\lambda_1, \lambda_2, ..., \lambda_n\}$ which are the roots of the determinant of the matrix polynomial $D_R$. This means finding $\lambda_i$ such that $P(\lambda_i) = 0$ where $P(s) = \det(D_R(s))$.

***Step 2:*** Sort the latent roots in terms of their real part.

***Step 3:*** Choose the latent roots $\lambda_i$ which are farthest to the imaginary axis. Then, Compute their corresponding right latent vectors $v$ using the formula $A(\lambda_i) v = o$.

***Step 4:*** Check if the computed latent vectors are linearly independent. If not, remove the chosen latent roots from the list and go to step 3.

***Step 5:*** Compute the solvent using the formula $R = V \Lambda V^{-1}$ where $\Lambda = \{\lambda_1, \lambda_2, ..., \lambda_m\}$ are the latent roots and $V = \{v_1, v_2, ..., v_m\}$ are the corresponding linearly independent latent vectors.

***Step 6:*** Compute $D_{MOR}(\lambda)$ such that $D_R(\lambda) = D_{MOR}(\lambda)(\lambda I - R)$. If the resulting system is not proper, compute also $N_{MOR}$ such that $N_R(\lambda) = N_{MOR}(\lambda)(\lambda I - R) + \Re$.

***Step 7:*** Form the new transfer function $G_{MOR}(s) = N_{MOR}(s) D_{MOR}(s)^{-1}$.

Notice that if the given system is in state space form, it can be transformed to a coprime matrix fraction. Also, instead of using right solvents, one can use left solvents in a dual manner.
This procedure is repeated until the error between the original and the resulting system is less than the given tolerance.
This method was tested on several systems. Although it worked for a limited number of cases, it did not in most cases. This is due mostly to the fact that the algorithm could not find linearly independent latent vectors that yield solvents that are insignificant. Since this method did not yield satisfying results, we introduce a more robust method in section 5.

# 4 Method based on dominant poles

Given a system in state space form and a complete set of right solvents, one can write the transfer function as an addition of subsystems. The complete set of solvents can be obtained using different methods [19, 21, 22]. In our implementation, the block power method was used [21].

First, the system is transformed into a controller canonical form:

$$\dot{X}_C = A_C X_C + B_C U$$
$$Y = C_C X_C \tag{20}$$

$A_C$ is of size $n \times n$, $B_C$ of size $n \times m$ and $C_C$ of size $p \times n$. The matrix D is discarded in (20) and from the following equations since it is not affected by the transformation.

Since the complete set of solvents is available, one can write the Vandermonde matrix:

$$V_R = \begin{bmatrix} I_m & I_m & \ldots & I_m \\ R_1 & R_2 & \ldots & R_r \\ \vdots & \vdots & \ldots & \vdots \\ R_1^{r-1} & R_2^{r-1} & \ldots & R_r^{r-1} \end{bmatrix} \tag{21}$$

where $R_i$ for i=1,2,…,r are the right solvents of the MIMO system.

Then, using the transformation:

$$X_C = V_R X_R \tag{22}$$
$$X_R = V_R^{-1} X_C \tag{23}$$

To get the state space equations we take the derivative:

$$\dot{X}_R = V_R^{-1}(A_C X_C + B_C U) \tag{24}$$

Replacing $X_C$ results in:

$$\dot{X}_R = (V_R^{-1} A_C V_R) X_R + (V_R^{-1} B_C) U \tag{25}$$

Replacing also $X_C$ for the output:

$$y = (C_C V_R) X_R \tag{26}$$

This results in the new matrices:

$$A_R = V_R^{-1} A_C V_R$$
$$B_R = V_R^{-1} B_C \tag{27}$$
$$C_R = C_C V_R$$

The state space equations will have the form:

$$\begin{bmatrix} \dot{X}_1 \\ \dot{X}_2 \\ \vdots \\ \dot{X}_r \end{bmatrix} = \begin{bmatrix} R_1 & 0 & \ldots & 0 \\ 0 & R_2 & \ldots & 0 \\ \vdots & \vdots & \ldots & \vdots \\ 0 & 0 & \ldots & R_r \end{bmatrix} \begin{bmatrix} X_1 \\ X_2 \\ \vdots \\ X_r \end{bmatrix} + \begin{bmatrix} B_1 \\ B_2 \\ \vdots \\ B_r \end{bmatrix} U$$

(28)

$$Y = \begin{bmatrix} C_1 & C_2 & \ldots & C_r \end{bmatrix} \begin{bmatrix} X_1 \\ X_2 \\ \vdots \\ X_r \end{bmatrix}$$

This results in a block decoupled transfer function of the form:

$$G(s) = C_1(sI - R_1)^{-1}B_1 + C_2(sI - R_2)^{-1}B_2 + \ldots + C_r(sI - R_r)^{-1}B_r \quad (29)$$

This form allows to eliminate directly the subsystems corresponding to specific solvents.

We know that some poles don't affect the system significantly and can be discarded without a noticeable effect on the system response. To determine which poles are essential, we use Subspace Accelerated MIMO Dominant Pole Algorithm (SAMDP) presented in [23]. This algorithm is fast, reliable and flexible. It computes the dominant poles one by one by selecting the most dominant approximation in every iteration.

SAMDP computes the dominant poles and corresponding residue matrices one by one by selecting the most dominant approximation in every iteration. This approach leads to a faster, more robust, and more flexible algorithm. To avoid repeated computation of the same dominant poles, a deflation strategy is used.

The algorithm is based on a modified accelerated Newton scheme. The problem is to compute dominant poles $\lambda_j$ and the corresponding right and left eigenvectors $x_j$ and $y_j$.

The method was developed for square and non-square transfer functions. For simplicity, we will consider only the square transfer function H(s) here. The other cases can be found in [23].

The dominant poles are those $s_i \in \mathbb{C}$ for which $\sigma_{\max}(H(s)) \to \infty$. For square transfer functions this is equivalent to finding $s_i \in \mathbb{C}$ for which $\lambda_{\max}(H^{-1}(s)) \to 0$.

Then as proven in [23], a dominant pole is computed iteratively as:

$$s_{k+1} = s_k - \frac{1}{\mu_{\min} v^* C^T (s_k I - A)^{-2} B u} \quad (30)$$

where $(\mu_{\min}, u, v) = (\mu_{\min}(s_k), u_{\min}(s_k), v_{\min}(s_k))$ is the eigentriplet of

$H^{-1}(s_k)$ corresponding to $\lambda_{min}(H^{-1}(s_k))$.

Then three strategies are used to make this algorithm able to compute more than one dominant pole: subspace acceleration, selection of most dominant approximation, and deflation. The subspace acceleration keeps the search spaces orthogonal using modified Gram Schmidt (MGS). The selection strategy concerns the selection of the new pole estimate. This algorithm is sensitive to this choice. In this algorithm the new pole is selected based on the largest residue norm. Finally, a deflation strategy is used to avoid computing already converged poles. For a real system, if a complex pole converged, its complex conjugate is also a pole.

The detailed algorithms and MATLAB code can be found in [23]. This algorithm needs an initial guess for the poles which can affect the result. The obtained dominant poles are then matched with their corresponding solvents. The k selected solvents are kept with their corresponding subsystems. All the other subsystems are discarded. This allows to eliminate multiple solvents at the same time. The whole model reduction procedure is summarized in figure 1.

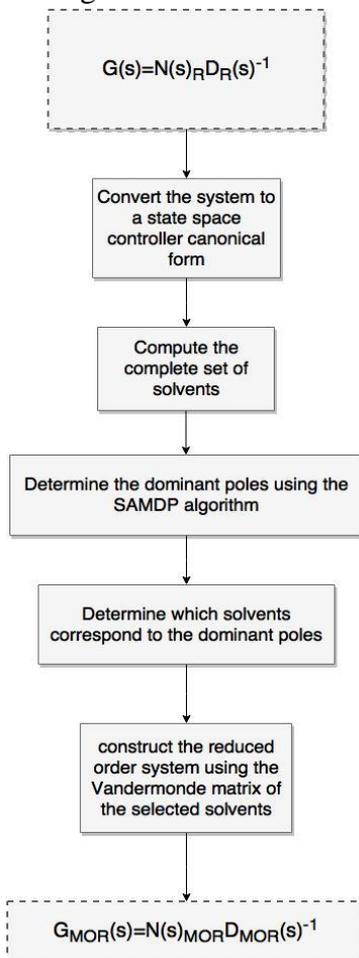

Fig 1. Block diagram of the model reduction method based on dominant poles.

Once the system with k solvents is obtained, one can try to further eliminate solvents one by one and verify if the error criterion between the given system and the new system is satisfied. If we cannot reduce the system further, one can turn to the eigenvalue level and try to eliminate the eigenvalues one by one in the last subsystem.

Notice that instead of using right solvents, left solvents can be used. The left solvents method is dual to the one presented here, this is why it is not presented.

The presented algorithm was implemented in MATLAB and tested on various examples. To illustrate its use, an example is presented.

## 5 Numerical Results

Electrical power networks can be quite complex. In general these systems require model reduction before being able to tackle analysis or design. Take for example the electric power network [24] of order 8 in state space form:

$$A_r = \begin{bmatrix} 0 & 0 & 1 & 0 & 0 & 0 & 0 & 0 \\ 0 & 0 & 0 & 1 & 0 & 0 & 0 & 0 \\ 0 & 0 & 0 & 0 & 1 & 0 & 0 & 0 \\ 0 & 0 & 0 & 0 & 0 & 1 & 0 & 0 \\ 0 & 0 & 0 & 0 & 0 & 0 & 1 & 0 \\ 0 & 0 & 0 & 0 & 0 & 0 & 0 & 1 \\ 7648 & 2011 & 4144 & 3043 & 547 & 331 & 12 & 22 \\ -24106 & -5888 & -13914 & -9446 & -2160 & -1124 & -104 & -80 \end{bmatrix} \quad B_r = \begin{bmatrix} 0 & 0 \\ 0 & 0 \\ 0 & 0 \\ 0 & 0 \\ 0 & 0 \\ 0 & 0 \\ 1 & 0 \\ 0 & 1 \end{bmatrix}$$

$$C_r = \begin{bmatrix} 99.7751 & 199.8833 & -14.9965 & -45.0149 & 3.0007 & -1.0003 & 0.1000 & -0.1000 \\ -100.2466 & 99.9340 & 60.0005 & 44.9846 & -19994 & 3.9998 & -0.2000 & 0.1000 \end{bmatrix} \quad (31)$$

The complete set of solvents of this system can be computed using the Power method. Their corresponding latent roots are also shown in Table 1.

Next, we need to determine which solvents are essential. We run the SAMDP algorithm to determine the dominant poles. These poles are shown in equation (32).

$$\begin{aligned} p_1 &= -6.1407 + 0.0000i \\ p_2 &= -4.6097 + 0.0000i \\ p_3 &= -10.5725 + 0.0000i \\ p_4 &= -0.7967 + 1.2075i \\ p_5 &= -0.7967 + 1.2075i \\ p_6 &= -0.7967 - 1.2075i \end{aligned} \quad (32)$$

As defined in [23], the dominant poles determine most of the system response. Therefore, eliminating the non-dominant poles does not affect drastically the system response. Comparing the dominant poles from $p_1$ to $p_6$ in equation(32), to the system poles represented by the eigenvalues of the complete set of solvents in table 1, we can see that dominant poles are present in solvents R1, R2 and R4. So R3 can be eliminated without affecting too much the system response. This comparison can be done automatically by iterating over the solvents and check if their eigenvalues include one of the dominant poles.

$$A_r = \begin{bmatrix} 0 & 1 & 0 & 0 & 0 & 0 \\ 0 & 0 & 1 & 0 & 0 & 0 \\ 4 & -529 & 13 & 137 & 88 & 19 \\ 0 & 0 & 0 & 0 & 1 & 0 \\ 0 & 0 & 0 & 0 & 0 & 1 \\ -145 & 1570 & -48 & -428 & -288 & -65 \end{bmatrix} \quad B_r = \begin{bmatrix} 0 & 0 \\ 0 & 0 \\ 1 & 0 \\ 0 & 0 \\ 0 & 0 \\ 0 & 1 \end{bmatrix}$$

$$C_r = \begin{bmatrix} -47 & 13 & 1 & 5 & 1 & 0 \\ -37 & -7 & 0 & 0 & 2 & 0 \end{bmatrix}$$

(33)

As it can be seen in (33), the resulting system has order 6.The resulting system has indeed almost the same response as the original system as shown in the comparison between the Bode plots of the original system and the resulting system in figure 2.

Notice that here are interested in the input output behavior of the system represented by the Bode plot. Other methods are being developed taking into account the preservation of the structure and the physical meaning of the system. These methods use Bond graphs and Port Hamiltonian representations.

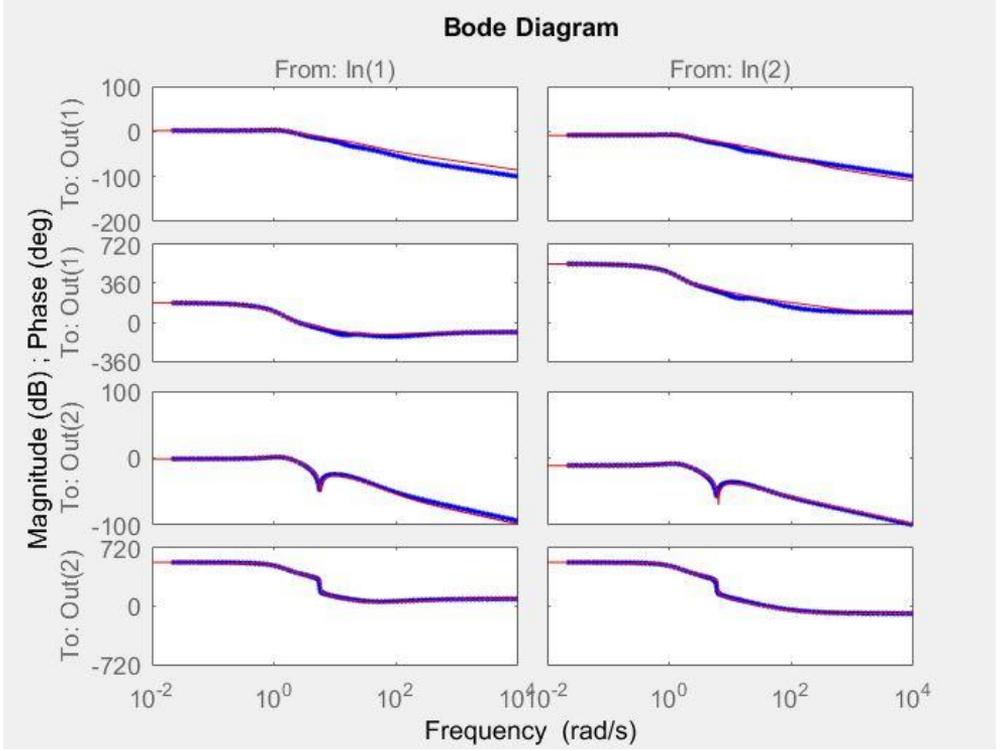

Fig 2. Bode plot of the given system (blue dots) and the resulting reduced system (red line)

In addition to the Bode plot, the original and reduced systems are compared quantitatively using the measure mentioned in [8]. A reduced model is considered to be close to the original one if the sum of the Hankel principal components of the neglected part of the system is small compared with the sum of the Hankel principal components of the whole system. That is the relative error (RE) shown in equation (34) should be less than a set threshold.

$$RE = \frac{\left(\sum_{i=1}^{r} \bar{\sigma}_i^4\right)^{1/2}}{\left(\sum_{j=1}^{n} \sigma_j^4\right)^{1/2}} \quad (34)$$

where $\bar{\sigma}_i$ are the Hankel principal components of the neglected part and $\sigma_j$ are the Hankel principal components of the of the whole system.

Because of the nature of the problem we set the threshold to be 0.01 . For this example the final relative error is RE=0.0021 . Other methods based on Krylov subspaces resulted in less error but the contributed method is an alternative method that results in a special canonical form. This also allows to reduce the storage needed to store the result as only the nontrivial value are stored in blocks. For example, the resulting B matrix does not need to be stored. Only the dimension of B and the size of its blocks are needed to recover it.

# 6 Discussion and conclusions

In this paper, we presented two model reduction methods for MIMO linear systems. Both methods are based on solvents or block roots. The advantage of using solvents stems from of the rich theory of matrix polynomials. Although model reduction has been researched extensively, methods based on matrix polynomials were not fully exploited. Since we are dealing with matrix polynomials, these methods are exclusively applied to MIMO systems.

The two contributed methods have their advantages and disadvantages with respect to other model reduction methods. Although the method based on the latent structure (presented in section 4) is not applicable to all systems because it is not always possible to construct solvents using this method, it offers the advantage that the solvents can be eliminated one by one without having to compute the complete set of solvents. Eliminating solvents one by one reduces also the computation time and the memory space used by the algorithm. Finally, the algorithm keeps the system in a matrix polynomial fraction form which allows for direct compensator design [25].

On the other hand, the method presented in section 5 requires the computation of the complete set of solvents first. However, this method allows to discard multiple solvents at the same time which reduces computation time. Also, this method allows to tune the system even further by eliminating solvents one by one and even poles one by one. This method is more suitable for pole placement or Eigenstructure placement design [26, 27] since the result can be easily written in state space form.

# 7 References

1. S. Gugercin and A. C. Antoulas, "A Survey of Model Reduction by Balanced Truncation and Some New Results," *International Journal of Control,* vol. 77, no. 8, pp. 748-766, 2004.
2. P. Benner, S. Gugercin, and K. Willcox, "A Survey of Projection-Based Model Reduction Methods for Parametric Dynamical Systems," *SIAM Review,* vol. 57, no. 4, pp. 483-531, 2015.
3. A. C. Antoulas, "An overview of approximation methods for large-scale dynamical systems," *Annual Reviews in Control,* vol. 29, no. 2, pp. 181-190, 2005.
4. M. Green and B. D. O. Anderson, "Model reduction by phase matching," *Mathematics of Control, Signals and Systems,* vol. 2, no. 3, pp. 221-263, 1989.
5. C. K. Chui, X. Li, and J. D. Ward, "System reduction via truncated Hankel matrices," *Mathematics of Control, Signals and Systems,* vol. 4, no. 2, pp. 161-175, 1991.
6. Y. Genin and A. Vandendorpe, "On the embedding of state space

realizations," *Mathematics of Control, Signals, and Systems,* vol. 19, no. 2, pp. 123-149, 2007.
7.  B. Bekhiti, A. Dahimene, B. Nail, and K. Hariche, "On The Order Reduction of MIMO Large Scale Systems Using Block-Roots of Matrix Polynomials," ed.
8.  H. Zabot and K. Hariche, "On solvents-based model reduction of MIMO systems," *International Journal of Systems Science,* vol. 28, no. 5, pp. 499-505, 1997.
9.  I. Gohberg, P. Lancaster, and L. Rodman, *Matrix Polynomials* (Classics in Applied Mathematics). Society for Industrial and Applied Mathematics, p. 423, 2009.
10. K. Hariche and E. D. Denman, "Interpolation theory and λ-matrices," *Journal of Mathematical Analysis and Applications,* vol. 143, no. 2, pp. 530-547, 1989.
11. K. Hariche and E. D. Denman, "On solvents and Lagrange interpolating λ-matrices," *Applied Mathematics and Computation,* vol. 25, no. 4, pp. 321-332, 1988.
12. M. Yaici and K. Hariche, "On solvents of matrix polynomials," *international journal of modeling and optimization,* vol. 4, no. 4, pp. 273-277, 2014.
13. W. H. Schilders, H. A. Van Der Vorst, and J. Rommes, *Model Order Reduction: Theory, Research Aspects and Applications* (The European Consortium for Mathematics in Industry). Springer-Verlag Berlin Heidelberg, 2008, pp. XI, 471.
14. L. Montier, T. Henneron, S. Clénet, and B. Goursaud, "Robust model order reduction of an electrical machine at startup through reduction error estimation," *International Journal of Numerical Modelling: Electronic Networks, Devices and Fields,* vol. 31, no. 2, 2018, Art. no. e2277.
15. S. S. Mohseni, M. J. Yazdanpanah, and A. Ranjbar N, "New strategies in model order reduction of trajectory piecewise-linear models," *International Journal of Numerical Modelling: Electronic Networks, Devices and Fields,* vol. 29, no. 4, pp. 707-725, 2016.
16. H. Wu and A. C. Cangellaris, "Krylov model order reduction of finite element approximations of electromagnetic devices with frequency-dependent material properties," *International Journal of Numerical Modelling: Electronic Networks, Devices and Fields,* vol. 20, no. 5, pp. 217-235, 2007.
17. P. J. Heres, D. Deschrijver, W. H. A. Schilders, and T. Dhaene, "Combining Krylov subspace methods and identification-based methods for model order reduction," *International Journal of Numerical Modelling: Electronic Networks, Devices and Fields,* vol. 20, no. 6, pp. 271-282, 2007.
18. M. V. Ugryumova, J. Rommes, and W. H. A. Schilders, "Error bounds for reduction of multi-port resistor networks," *International Journal of Numerical Modelling: Electronic Networks, Devices and Fields,* vol. 26, no. 5, pp. 464-477, 2013.
19. J. S. H. Tsai, C. M. Chen, and L. S. Shieh, "A computer-aided method for solvents and spectral factors of matrix polynomials," *Applied Mathematics *and Computation,* vol. 47, no. 2, pp. 211-235, 1992.
20. P. Benner, T. Breiten, and L. Feng, "Matrix Equations and Model Reduction," in *Matrix Functions and Matrix Equations*, vol. Vol 19(Series in Contemporary Applied Mathematics, no. Vol 19): Co-published with HEP, pp. 50-75, 2015.
21. J. S. H. Tsai, L. S. Shieh, and T. T. C. Shen, "Block power method for computing solvents and spectral factors of matrix polynomials," *Computers & Mathematics with Applications,* vol. 16, no. 9, pp. 683-699, 1988.
22. L. S. Shieh, Y. T. Tsay, and N. P. Coleman, "Algorithms for solvents and spectral factors of matrix polynomials," *International Journal of Systems Science,* vol. 12, no. 11, pp. 1303-1316, 1981.
23. J. Rommes and N. Martins, "Efficient Computation of Multivariable Transfer Function Dominant Poles Using Subspace Acceleration," *IEEE Transactions on Power Systems,* vol. 21, no. 4, pp. 1471-1483, 2006.
24. L. Fortuna, G. Nunnari, and A. Gallo, *Model Order Reduction Techniques with Applications in Electrical Engineering*. Springer-Verlag London, 1992.
25. M. Yaici, K. Hariche, and T. Clarke, "Helicopter flight control compensator design," *AIP Conference Proceedings,* vol. 1798, no. 1, p. 020174, 2017.
26. M. Yaici and K. Hariche, "On eigenstructure assignment using block poles placement," *European Journal of Control,* vol. 20, no. 5, pp. 217-226, 2014.
27. B. Bekhiti, A. Dahimene, B. Nail, K. Hariche, and A. Hamadouche, "On Block roots of matrix polynomials based MIMO control system design," in *2015 4th International Conference on Electrical Engineering (ICEE)*, 2015.

*Table 1. The complete set of solvents and their corresponding eigenvalues*

| Name | Solvent | Eigenvalues of the solvent |
|------|---------|----------------------------|
| R1 | $\begin{bmatrix} 0 & 0.8000 \\ -2.6000 & -1.6000 \end{bmatrix}$ | • -0.8000+1.2000i <br> • -0.8000-1.2000i |
| R2 | $\begin{bmatrix} -10.4500 & -28.5000 \\ 0.9500 & 0 \end{bmatrix}$ | • -5.7000 <br> • -4.7500 |
| R3 | $\begin{bmatrix} -1.5000 & 3.0000 \\ -56.000 & -15.000 \end{bmatrix}$ | • -8.2500+11.0651i <br> • -8.2500-11.0651i |
| R4 | $\begin{bmatrix} -10.0000 & 10.0000 \\ -2.0000 & -30.0000 \end{bmatrix}$ | • -11.0557 <br> • -28.9443 |